\documentclass[rsi,twocolumn,groupedaddress,showpacs]{revtex4}
\usepackage[dvipdfmx,colorlinks=true,citecolor=blue,linkcolor=blue,pdfhighlight =/O]{hyperref}
\usepackage{graphicx}
\usepackage{amsmath}
\usepackage{color}
\usepackage{soul}
\begin{document}
\title{A subKelvin scanning probe microscope for the electronic spectroscopy of an individual nano-device}
\author{T.~Quaglio, F.~Dahlem, S.~Martin, A.~G\'erardin, C.~B.~Winkelmann, and H.~Courtois}
\affiliation{
Institut N\'eel, CNRS, Grenoble INP and Universit\'e Joseph Fourier, 25 rue des Martyrs, BP 166, 38042 Grenoble, France.}
\date{\today}
\begin{abstract}
We present a combined scanning force and tunneling microscope working in a dilution refrigerator that is optimized for the study of individual electronic nano-devices. This apparatus is equipped with commercial piezo-electric positioners enabling the displacement of a sample below the probe over several hundred microns at very low temperature, without excessive heating. Atomic force microscopy based on a tuning fork resonator probe is used for cryogenic precise alignment of the tip with an individual device. We demonstrate the local tunneling spectroscopy of a hybrid Josephson junction as a function of its current bias.
\end{abstract}
\maketitle

\section{Introduction}

Scanning Tunneling Spectroscopy (STS) has been extensively used in the context of condensed matter physics for the local study of fundamental electronic properties including superconductivity, magnetism, Schockley surface states. Operation at very low temperature \cite{RSI-Moussy,RSI-Suderow,PRL-LeSueur,RSI-Stroscio} improves the spectroscopic energy resolution, with values down to 10 $\mu$eV being demonstrated \cite{RSI-LeSueur}. Recently, new cryogenic scanning tunneling microscopes enabling tunneling studies under out-of-equilibrium conditions have been developed~\cite{APL-Senzier,RSI-Maldonado}. The next step is to build instruments for the scanning tunneling spectroscopy of an individual nano-scale device. As electron tunneling is inoperative on an insulator, locating at the submicron scale a device sitting on an insulating substrate requires the use of Atomic Force Microscopy (AFM). AFM and STS can be combined on a single probe by using a quartz resonator, which remains stiff at low frequency, functionnalized with a metallic tip allowing to collect a tunnel current \cite{APL-Senzier,RSI-Smit}. As the initial tip approach is usually realized at a random distance from the sample, one needs to realize large lateral displacements at low temperature. Due to the difficulty of combining these different requirements, the tunneling spectroscopy of an individual nano-device, that can be current-biased, has to the best of our knowledge not yet been demonstrated.

In this paper, we demonstrate the local spectroscopy of an individual current-biased superconducting nano-device in the subKelvin temperature range. In this scope, we have developed a home-made combined AFM/STM based on a quartz tuning fork probe and equipped with commercial piezo-electric positioners. We describe the very low temperature operation, calibration and thermal behavior of these positioners. An optimized process for in-situ single nano-object localization is presented. As a proof of concept, we discuss a local spectroscopy experiment on a Superconductor-Normal metal-Superconductor (SNS) Josephson junction under current bias. 

\section{The cryogenic microscope}

\begin{figure}[t]
\includegraphics[width=\columnwidth]{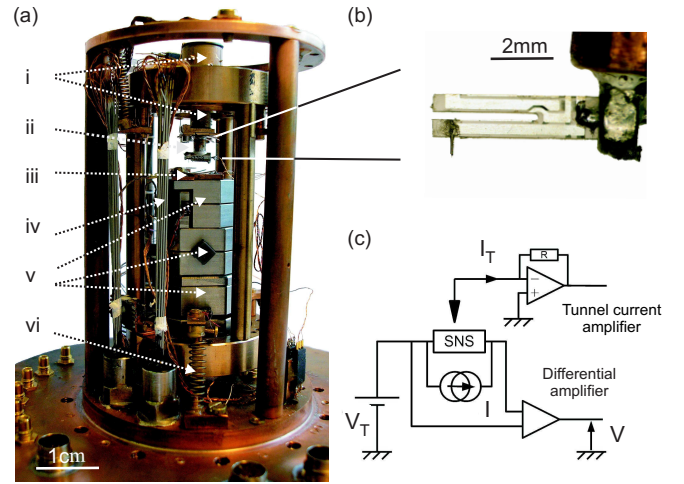}
\caption{
a) View of the cryogenic microscope head suspended on a 10 cm-high copper frame: (i) piezoelectric scanner tubes, (ii) AFM/STM probe, (iii) sample holder, (iv) micro-coax filters, (v) Attocube piezo-electric positioners and (vi) mechanical isolation springs. 
b) Zoom-in picture on the AFM/STM probe showing the tungsten tip glued by silver epoxy on a quartz tuning fork, here without its shielding box. 
c) Schematic of the circuit connecting the SNS sample, enabling to simultaneously perform $I(V)$ transport experiments and local tunnel spectroscopies $I_T(V_T)$.
\label{fig1_head}}
\end{figure} 

Fig.~\ref{fig1_head}a shows the overview of the microscope head that is suspended on a copper frame through six springs. It includes an arrangement of piezo-electric tubes at the top and a set of piezo-electric positioners at the bottom. 

The microscope head top part (number i in Fig.~\ref{fig1_head}a) holds the tip. It is made of two concentric EBL2 piezo-electric tubes connected in series but in opposite directions. The outer tube is used to adjust the tip-sample distance and, in the AFM mode, to mechanically excite the tuning fork. The inner tube controls the lateral scanning of the probe over the sample surface. The tube lengths are 1.25 and 2.5 cm, diameters are 1.25 and 0.62 cm, wall thicknesses are 1 and 0.5 mm, respectively. For a 220~V applied voltage on the inner tube and at low temperature, the maximum expansion in Z direction is about $\rm{1.2~\mu m}$ and the X-Y scanning area is about $\rm{6.2 \times 6.2~\mu m^2}$. We took special care when gluing the piezoelectric tubes to their metallic supporting parts, since even a small electric leakage to the ground may, under high voltage bias, create an important Joule heat, which could heat significantly the cryogenic system. 

A set of three piezo-electric positioners from Attocube \cite{Attocube} (number v in Fig.~\ref{fig1_head}a) is used to move the sample under the probe for coarse positioning in the three space directions. From bottom to top, one finds the Z (vertical) direction positioner, the Y and the X (horizontal) direction ones. The operation of this system will be described in detail in section III. The sample holder (number iii in Fig.~\ref{fig1_head}a) is made from a thermal clad plate that is screwed to the upper piezo-electric positioner. The sample is fixed on the sample holder with Apiezon N vacuum grease, which offers a good thermal link and leaves the possibility to easily remove the sample. Sample on-chip electrical contacts are connected to millimeter-sized copper pads defined on the sample holder through micro-bonded 30 $\mu$m-diameter Al wires.

During low temperature operation, the whole microscope is enclosed in a hermetic copper chamber that is mounted on the cold plate of an inverted dilution refrigerator. This refrigerator called Sionludi \cite{Sionludi} is compact and thus stiff. It presents a large available volume of about $1~dm^3$ at very low temperature that is most suitable for the operation of a scanning probe microscope. Its cooling power is about $\rm{30~\mu W}$ at 100~mK~\cite{RSI-Moussy}. The cryostat sits on on a set of four damped legs that filter mechanical vibrations from the ground floor with a cut-off at about 2 Hz.

Special care is taken for proper thermalization of the whole system. A $\rm{10^{-1}~mbar}$ of Helium exchange gas inserted in the microscope chamber before cool-down improves thermalization of the whole system down to about 4.2 K. In order to ensure a proper thermalization and a good electrical ground, every stage of the positioners tower is connected, through a copper wire, to the base of the microscope frame. In this way, the temperature on the sample stage reaches a base temperature below 100~mK. The copper chamber also protects the measurement set-up against electromagnetic radiation. In order to ensure proper electronic thermalization, every electrical line is filtered at the microscope chamber input through lossy micro-coaxial cables \cite{JAP-Glattli} that are thermally anchored to the refrigerator cold plate. 

\section{The AFM-STM probe}

The local probe (Fig.~\ref{fig1_head}b and number ii in Fig.~\ref{fig1_head}a) is a quartz tuning fork of 32768~Hz bare resonance frequency with an etched tungsten tip glued on it. The tuning fork oscillation is electrically detected. Compared to an optical detection, this approach is well adapted to very low temperatures: no optical window is necessary and dissipation is less than 1~nW \cite{APL-Karrai,ASS-Rychen}. The original metallic casing of the tuning fork is usually kept in order to shield the tip from electromagnetic noise and to simplify its gluing to the probe holder. In this case, only a little hole is drilled into it to access the fork electrode on which the tip will be glued.  

A sharp and stable tip is crucial to perform AFM on small structures with a large relief. The tip is produced by a 2 minutes electrochemical etch of a 150~$\mu$m-diameter tungsten wire in a 4 mol/l KOH~solution \cite{RSI-Kulawik}. The typical geometry of the tip apex is conical with an apex curvature of about 20~nm. The tip wire is then glued with silver epoxy \cite{Epotek} on one electrode of the tuning fork. This silver epoxy guarantees both an ohmic contact and a very stiff gluing. A relatively large mass of silver epoxy is used, so that a resonance frequency of about 30 kHz and a quality factor Q of about 7000 is obtained in air at room temperature. This routinely gives a quality factor of about $10^{5}$ at low temperature under cryogenic vacuum. Obtaining larger Q values is possible but this was found to offer no useful increase of sensitivity while bringing oscillation instabilities during imaging. As a tip, we have also used commercial AFM cantilever-tips coated with 25~nm of platinum-iridium. By using a very thin layer of silver epoxy and a micro manipulator, the cantilever can easily be attached by capillarity to the end of the tuning fork and broken at the desired position, thus leaving a pyramidal and metal-coated tip glued on the tuning fork prong. The latter method can bring a more stable tip geometry at the expense of the risk of loosing tunnel contact if the tip metallization gets damaged. This probe is finally glued again with silver epoxy on a copper plate, which is screwed to the piezo-electric scanner tube. This configuration offers proper mechanical coupling to the excited piezo-electric tube, which avoids any unwanted resonance frequency drift.

The control of the tuning fork oscillation is made through a Phase-Locked Loop included in a SPM electronics \cite{Nanonis} that is also used to control the whole system, including coarse approach and positioning, imaging and spectroscopy measurements. 
 
\section{Coarse positioning operation}

\begin{figure}
\includegraphics[width=0.95\columnwidth]{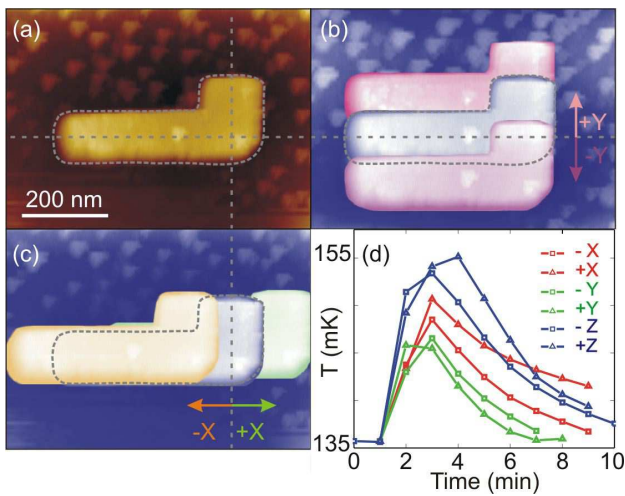}
\caption{Operation of the coarse positioning system at a bath temperature of 90 mK.
a) AFM topographical image of a tracking symbol. 
b) Superposed images after a move in - Y and + Y directions, with 8 pulses of 45 V amplitude. 
c) Superposed images after a move - X and + X movements, with 3 pulses of 45 V amplitude. 
d) Temperature evolution of the refrigerator cold plate due to a X, Y or Z piezo-electric positioners displacement of 10 steps at a 45 V bias amplitude.
\label{fig2_calibration}}
\end{figure}  

Our piezo-electric positioners \cite{Attocube} operation is based on a slip-stick motion. A sawtooth voltage with an amplitude up to 70 V is applied with a repetition rate of 1~ms. The decay time of the sawtooth fast slope is about $\rm{1~\mu s}$. Due to the large capacitance of the piezo-electric actuator, a small line resistance is required in order to maintain a small RC time and thus an efficient transfer of the signal waveform to the piezo-electric actuator. Between the room temperature stage and the cold plate, we use a 1-meter cable of 12 constantan wires of 100~$\Omega$ resistance. At low temperature, the piezo-electric actuator capacitance is only 22 or 100~nF for the X, Y or Z positioners respectively, so that 2 or 4 wires in parallel are enough to maintain an efficient operation. In order to reduce the number of required wires, the reference electrode of every motor is put to ground at the microscope copper frame. At room temperature, the larger capacitance of the piezo-electric actuators leads us to shunt the main wiring described above with a copper wires cable, which is used for the first rough adjustment of the tip above the device in air. Before any transport or tunneling spectroscopy measurement, every positioner is connected to the electrical ground in order to avoid electrical noise. 

Sample displacements have been calibrated using a substrate pattern made by electron beam lithography. Fig.~\ref{fig2_calibration}a illustrates a calibration sequence performed below 100~mK on one single symbol having a well-defined size. For X-translation (Fig.~\ref{fig2_calibration}b) and Y-translation (Fig.~\ref{fig2_calibration}c), only few pulses of 45~V are needed to move the sample on a 100 nm scale. Back and forth movements are in average reproducible at this scale. For the same translation, less pulses are needed for the X-stage that is at the top of the piezo-electric tables set. The Z-movement calibration (not shown here) was done via the distance variation of the tunnel contact position adjusted by the displacement of the calibrated piezo-electric scanner tube. Little difference between the top and down z-movements was found. In general, the main obstacle to displacements is given by the positioner electrical wires, which have to be positioned carefully around the microscope head. 

Fig.~\ref{fig2_calibration}d shows the thermal response of the dilution refrigerator mixing chamber after a series of ten steps of each positioner with a 45 V bias amplitude. Starting from temperature of 135 mK, the temperature increases of only 10 to 20 mK, whereas the temperature of the sample stage jumps to about 500 mK (not shown here). A few minutes are needed for the system to recover is base temperature after the sample move. Furthermore, it is possible to quickly move the sample over 10 micrometers by several hundred steps (of 45 V amplitude), while keeping the refrigerator stable at a temperature below about 1 K.

These experimental observations are compatible with basic thermal considerations. The positioners' nominal friction force of 5~N, a step size of about 30 nm and a displacement rate of 1~kHz determine a mechanical power of about $150~\mu W$ during movement. At a 45 V bias, the charging energy $C U^2/2$ stored in the 22 nF capacitance of the X or Y positioner is 560 $\mu J$. Under voltage pulses at a frequency $f$, the heat generation is $4 \pi\ C U^2 f \tan \delta$, where the loss angle $\delta$ is about 1 degree at low temperature~\cite{Bodefeld}. This gives a value of about $780~\mu W$. The total load of about a milliWatt is compatible with the estimated cooling power at 500 mK. Some extra heat dissipation can also originate from Ohmic losses in the electrical wires used to bias the piezoelectric positioners.
 
\section{Single device localization method}
 
\begin{figure}[t]
\includegraphics[width=0.95\columnwidth]{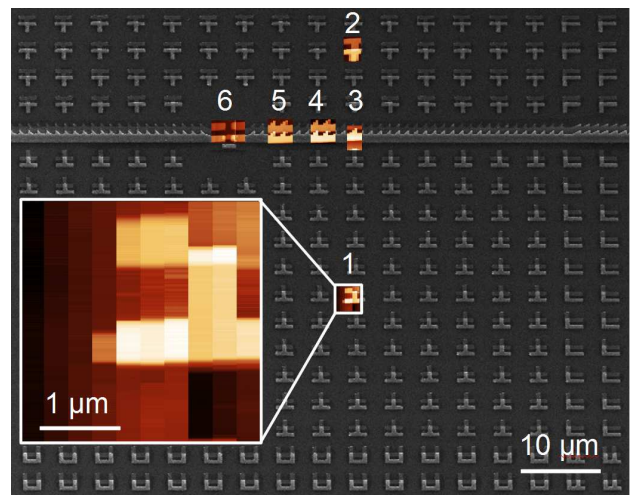}
\caption{
Scanning electron microscopy image showing the pattern of different symbols used to localize by AFM the sample at low temperature. Added on top of this image, the AFM pictures and associated numbers represent the different steps of a typical localization procedure performed at 90~mK. On the left: zoom on one of the low resolution image recorded during this procedure.
\label{fig3_pattern}}
\end{figure}  

A prerequisite for performing a local spectroscopy of an individual nano-device is to align the scanning probe with the device. For this purpose, the probe is first roughly aligned with the device at room temperature, using a binocular microscope. This is easily feasible due to the open design of the microscope head and the tip reflection visibility on the silicon substrate. While cooling down the system, the tip drifts by approximately $\rm{50~\mu m}$ due to the different thermal coefficients of the microscope components. We have therefore developed a reliable method for a precise tip alignment at low temperature, in practice at a temperature of about 1~K when thermal contractions are complete and the He pressure inside the chamber is low enough to avoid any sparkling. The method has been optimized for a minimum AFM imaging time, which is essential for saving the tip apex integrity for tunneling experiments. 

In this scope, the sample chip was previously covered with different kinds of symbols of micrometer size over a square area of $150 \times 150~\mu m^2$ centered on the nano-device of interest. The pattern design was defined such that one full symbol is always present within the low temperature scanning range. The patterned region is divided into 10 different zones, the $50 \times 50~\mu m^2 $ central zone being itself divided into 10 different smaller zones. Moreover, the patterns array is symmetric with respect to the two current leads. Each zone features a different symbol geometry that can be easily recognized in low-resolution AFM images. The individual pattern geometry is comb-shaped with a variable number of teeth. Typically, after 10 lines of scan with an inter-distance of 500~nm or 100~nm, the symbol recognition is achieved. Using the piezo-electric tables calibration, the tip is aligned above the nano-device by dichotomy after a maximum of ten AFM images. Fig.~\ref{fig3_pattern} shows a localization process case achieved in six steps, meaning that in this case only 60 scan lines in total were needed to find the nano-device. Once the device is located, the tip is moved over the metallic device, the tuning fork is let static so that electron tunneling spectroscopy can be performed. 

\section{Local spectroscopy on a current-biased proximity Josephson junction}

\begin{figure}[t]
\includegraphics[width=.95\columnwidth]{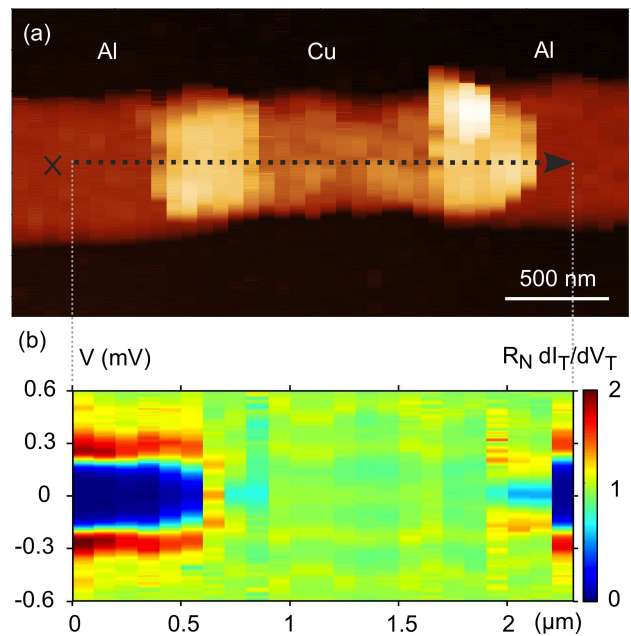}
\caption{
Typical measurements obtained with our cryogenic AFM/STM on a hybrid Josephson junction at 90~mK. 
a) AFM image of the Al/Cu/Al junction. The cross indicates where spectroscopies in out-of-equilibrium conditions were realized, see Fig. \ref{fig5_IV}b.
b) Color plot of local spectroscopic data $dI_T/dV_T(V_T)$ performed every 100~nm along the dotted arrow with a tunnel resistance of 10~$\rm{M \Omega}$.
\label{fig4_dIdV}}
\end{figure}

We have used the instrument and the method described above for the study of Superconductor-Normal metal-Superconductor (SNS) Josephson junctions. Samples were realized by shadow evaporation technique under ultra high vacuum on a Si wafer covered with $\rm{300~nm}$-thick $\rm{SiO_2}$. A Josephson junction is made of a $\rm{1.2~\mu m}$-long, $\rm{0.3~\mu m}$-wide and $\rm{35~nm}$ thick copper island bridging two $\rm{35~nm}$-thick aluminum leads. The two Al/Cu overlapping regions have a length about $\rm{0.2~\mu m}$ each. Low-power ultrasound and oxygen plasma cleaning were performed at the end of the lift-off process in order to ensure a clean surface. These precautions are helpful since traces of resist can be gathered by the tip during AFM measurements and affect the quality of both AFM images and tunnel spectroscopies. The AFM image in Fig.~\ref{fig4_dIdV}a shows the geometry and structure of the SNS junction.

\begin{figure}[t]
\includegraphics[width=0.95\columnwidth]{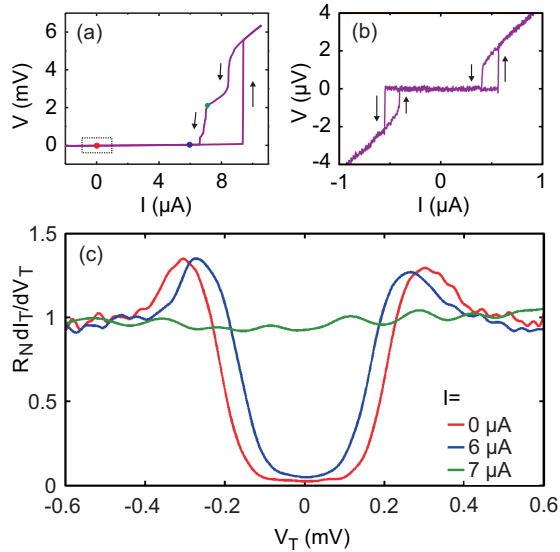}
\caption{ 
Typical measurements obtained with our cryogenic AFM/STM on a hybrid Josephson junction at 90~mK.
a) and b) Current-voltage characteristics $I(V)$ of the junction measured as described in Fig.~\ref{fig1_head}c and over two different bias current ranges. Some hysteresis is observed.
c) Local spectroscopic data $dI_T/dV_T(V_T)$ measured on one aluminum lead, see Fig. \ref{fig4_dIdV}, while current-biasing the sample close to the local critical current of the lead. 
\label{fig5_IV}}
\end{figure}

Fig.~\ref{fig5_IV}a shows a typical current-voltage characteristics $I(V)$ obtained at 90~mK on one of our samples. As expected, the data show a superconducting branch with a critical current of about 600 nA. At higher bias, the voltage reaches an Ohmic behavior with a slope given by the normal-state resistance of the junction of 4.5 $\Omega$. We observe an hysteretic behavior that is characteristic of heating effects in Josephson junctions with a large critical current~\cite{PRL-Courtois}. At a larger bias range, see Fig. \ref{fig5_IV}b, we observe the transition of the Al lead from the superconducting state to the resistive state. Again, this transition shows thermal hysteresis.

Local tunneling spectroscopies of the equilibrium local density of states (LDOS) were performed at various locations on the sample. The tunnel resistance between the tip and the device surface was fixed to 10~$\rm{M \Omega}$ and the tunnel current $I_T$ was measured when sweeping the voltage between tip and sample $V_T$. The data $I_T(V_T)$ was afterwards numerically differentiated as the tunnel differential conductance $dI_T/dV_T(V_T)$ gives a measure of the local electronic density of states, smeared by temperature. Fig.~\ref{fig4_dIdV}b 2D-color plot shows a set of differential conductance spectroscopies performed every 100~nm along the junction length. We can clearly see how the superconducting gap in every Aluminum lead (deep blue region) decreases in width when approaching the Al/Cu overlap region and finally vanishes in the copper metal part. The spatial evolution of the measured spectra along the X-axis was shown to be reproducible with a fidelity of at worst 50~nm even when the tunnel contact was established through the oxide covering an aluminum lead.

Tunneling spectroscopies measured on the superconducting Al leads are shown in Fig.~\ref{fig5_IV}c. The spectrum measured at zero bias current can be fitted assuming the Bardeen, Cooper and Schrieffer (BCS) density of states expression with a superconducting gap of 240~$\mu eV$ and an electronic temperature of 270~mK for the tip. This increased temperature compared to the measured sample temperature could be understood as due to some residual electronic noise in the set-up.

In the normal metal bridge of a SNS junction, theory predicts a density of states with a mini-gap given by 3.1 times the Thouless energy $E_{Th}=\hbar D/L^2$ \cite{JLTP-Pannetier,PRL-LeSueur}, where D is the electronic diffusion constant and L is the junction length. Assuming a typical value D = 70~cm$^2$ for Cu, taking L = 1 $\mu$m, we obtain an expected mini-gap value of 14~$\mu eV$, which is not observed. However, the thermal broadening due to the effective tip temperature $\rm{2k_B T}$=46~$\mu eV$ (at T = 270 mK) also determines the energetic resolution of our measurement and significantly smears smaller structures. The copper oxide at the surface may also destroy very locally the quantum correlations present inside the normal metal bridge.

Finally, we have investigated the effect of a current bias on the local spectroscopies on the normal metal bridge as well as on the superconducting electrodes region. Fig.~\ref{fig5_IV}c shows a series of local spectroscopies realized on one of the Al leads close to the junction, see Fig. \ref{fig4_dIdV}. No sizeable effect is detected until a current of about 6~$\mu$A, where the heating from the normal island becomes enough to affect locally superconductivity in the closely Al lead. At a bias current of 6~$\mu$A, the superconducting gap decreases from 240~$\mu eV$ at equilibrium down to about 200~$\mu eV$. For a current of 7~$\mu A$, the aluminum leads turn fully normal. The retrapping current of the superconducting wire measured in transport is actually 6.5~$\mu A$, see Fig. \ref{fig5_IV}a . This last experiment demonstrates the possibility to perform the local electronic spectroscopy of a quantum device under actual operation.

\section{Conclusion}

We have set up and characterized a home-made scanning probe microscope working in a dilution refrigerator that combines scanning force microscopy and tunneling spectroscopy, by using a tuning fork resonator functionnalized with a metallic tip. This system has the specificity to be equipped with commercial piezo-electric positioners whose operation is demonstrated to be compatible with subKelvin temperatures. We have developed a protocol based on a specially patterned substrate to position the tip over a single object at very low temperature. Thanks to this experimental setup, we succeeded in realizing the scanning tunneling spectroscopy of a nano-device biased with an electrical current.

\section{Acknowledgment}

The authors thank S.~Rajauria, L. M. A.~Pascal, T.~Crozes, A.~Fay, R.~Stomp, A. Schmalz and S. Falk for discussions. Samples have been fabricated at Nanofab facility at Institut N\'eel. We acknowledge the support from Grenoble Nanoscience Foundation and MicroKelvin, the EU FRP7 low temperature infrastructure grant 228464.

\end{document}